\documentclass[12pt,preprint]{aastex}

\shorttitle{HR 2582 Fundamental Parameters}
\shortauthors{Baines et al.}

\begin{document}

\title{Characterization of the Red Giant HR 2582 Using \\ the CHARA Array}

\author{Ellyn K. Baines}
\affil{Remote Sensing Division, Naval Research Laboratory, 4555 Overlook Avenue SW, \\ Washington, DC 20375}
\email{ellyn.baines@nrl.navy.mil}

\author{Harold A. McAlister, Theo A. ten Brummelaar, Nils~H.~Turner, Judit Sturmann, \\ Laszlo Sturmann, Christopher D. Farrington, Norm Vargas}
\affil{Center for High Angular Resolution Astronomy, Georgia State University, P.O. Box 3969, \\ Atlanta, GA 30302-3969}

\author{Gerard T. van Belle}
\affil{Lowell Observatory, Flagstaff, AZ 86001} 

\author{Stephen T. Ridgway}
\affil{National Optical Astronomy Observatory, P.O. Box 26732, Tucson, AZ 85726-6732} 

\begin{abstract}

We present the fundamental parameters of HR 2582, a high-mass red giant star whose evolutionary state is a mystery. We used the CHARA Array interferometer to directly measure the star's limb-darkened angular diameter (1.006$\pm$0.020 mas) and combined our measurement with parallax and photometry from the literature to calculate its physical radius (35.76$\pm$5.31 $R_\odot$), luminosity (517.8$\pm$17.5 $L_\odot$), bolometric flux (14.8$\pm$0.5 $\times$ 10$^{-8}$ erg s$^{-1}$ cm$^{-2}$) and effective temperature (4577$\pm$60 K). We then determined the star's mass (5.6$\pm$1.7 $M_\odot$) using our new values with stellar oscillation results from Baudin et al. Finally, using the Yonsei-Yale evolutionary models, we estimated HR 2582's age to be 165$^{+20}_{-15}$ Myr. While our measurements do not provide the precision required to definitively state where the star is in its evolution, it remains an excellent test case for evaluating stellar interior models.

\end{abstract}


\section{Introduction}

The red giant star HR 2582 (HD 50890, HIP 33243) was part of a study by \citet{2009AandA...506..465H}, who used the \emph{CoRoT} satellite \citep{2009AandA...506..411A} to observe G and K giant stars with solar-like oscillations. The distribution of these pulsating stars' seismic parameters indicted they belonged to the ``red clump'' of low-mass, post-flash core-He-burning, evolved stars \citep{2009AandA...503L..21M}. When the effective temperature and luminosity of these stars are not well characterized, it is difficult to determine their masses, ages, and radii \citep{2010AandA...509A..77K}. 

HR 2582 was of particular interest because its mass could be inferred using other data besides asteroseismology. \citet[][hereafter B12]{2012AandA...538A..73B} used spectroscopic observations to determine the following parameters: rotational velocity $v \sin i$ (10$\pm$2 km s$^{-1}$), effective temperature $T_{\rm eff}$ (4665$\pm$200 K), surface gravity log $g$ (1.4$\pm$0.3 cm s$^{\rm -2}$), metallicity [Fe/H] (-0.18$\pm$0.14), luminosity $L$ (log $L$ = 2.70$\pm$0.15 $L_\odot$), and finally a radius $R$ (34$\pm$8 $R_\odot$) using the Stefan-Boltzmann law.

HR 2582 was observed for 55 days using \emph{CoRoT} during its first science run. B12 found evidence for solar-like oscillations at low frequencies (between 10 and 20 $\mu$Hz) with a spacing of 1.7$\pm$0.1 $\mu$Hz between consecutive radial orders and noted that only radial modes are clearly visible in the data. They discovered an excess of power in the power density spectrum at $\nu_{\rm max} = 15 \pm 1 \mu$Hz and determined the star's mass using
\begin{equation}
\frac{\nu_{\rm max}}{\nu_{\rm max \odot}} \approx \frac{\frac{M}{M_\odot}}{(R/R_\odot)^2 \sqrt{\frac{T_{\rm eff}}{T_{\rm eff \odot}}}}.
\end{equation} 
Their value was $5.2 \pm 2.9$ $M_\odot$, which indicates HR 2582 is more massive than the stars in the red clump group described by Hekker et al. and Miglio et al. and implies rapid evolution. 

These results provide insights on the internal workings during the final evolutionary stages when the star is burning hydrogen in a shell, is burning its central helium, or is in the last stage of He-shell burning. While the star can be placed on the Hertzsprung-Russell diagram with relative precision, it is not sufficient to distinguish between the three evolutionary stages. Still, the results provide strong constraints on stellar interior models and are a good test case for those models (B12).

The advantage interferometry brings to HR 2582 is the ability to directly measure the angular diameter of the star instead of inferring its parameters using indirect methods. Then $R$ is determined using our angular diameter plus the distance to the star known from its parallax, and $T_{\rm eff}$ is calculated. We combine our results with those from stellar oscillation frequencies to more completely understand the star and determine its mass. Section 2 details our observing procedure; Section 3 discusses the visibility measurements and how stellar parameters were calculated, including angular diameter, radius, luminosity, and temperature; Section 4 explores the physical implications of the new measurements; and Section 5 summarizes our findings.

\section{Interferometric Observations}
We observed HR 2582 using the Center for High Angular Resolution Astronomy (CHARA) Array on 2012 December 12. The CHARA Array is a six-element optical-infrared interferometer located on Mount Wilson, California \citep{2005ApJ...628..453T}. We used the Classic beam combiner in the $K'$-band (2.13~$\mu$m) with the 279 m S1-W1 baseline.\footnote{The three arms of the CHARA Array are denoted by their cardinal directions: ``S'', ``E'', and ``W'' are south, east, and west, respectively. Each arm bears two telescopes, numbered ``1'' for the telescope farthest from the beam combining laboratory and ``2'' for the telescope closer to the lab. The ``baseline'' is the distance between the telescopes.} For a full description of the instrument, and the observing procedure and data reduction process used here, see \citet{2005ApJ...628..453T} and \citet{2005ApJ...628..439M}. 

When observing using an interferometer, selecting appropriate calibrator stars is extremely important because they are the standard against which we measure the scientific target. We used two calibrators, HD 46487 and HD 49434, which are both unresolved, single stars that acted as point sources. Because the stars' angular diameters are so small, uncertainties in their apparent sizes did not affect the target's diameter calculation as much as if they had a comparable angular size. We interleaved calibrator and target star observations so that every target was flanked by calibrator observations made as close in time as possible, which allowed us to convert instrumental target and calibrator visibilities to calibrated visibilities for the target. 

\begin{figure}[h]
\includegraphics[width=1.0\textwidth]{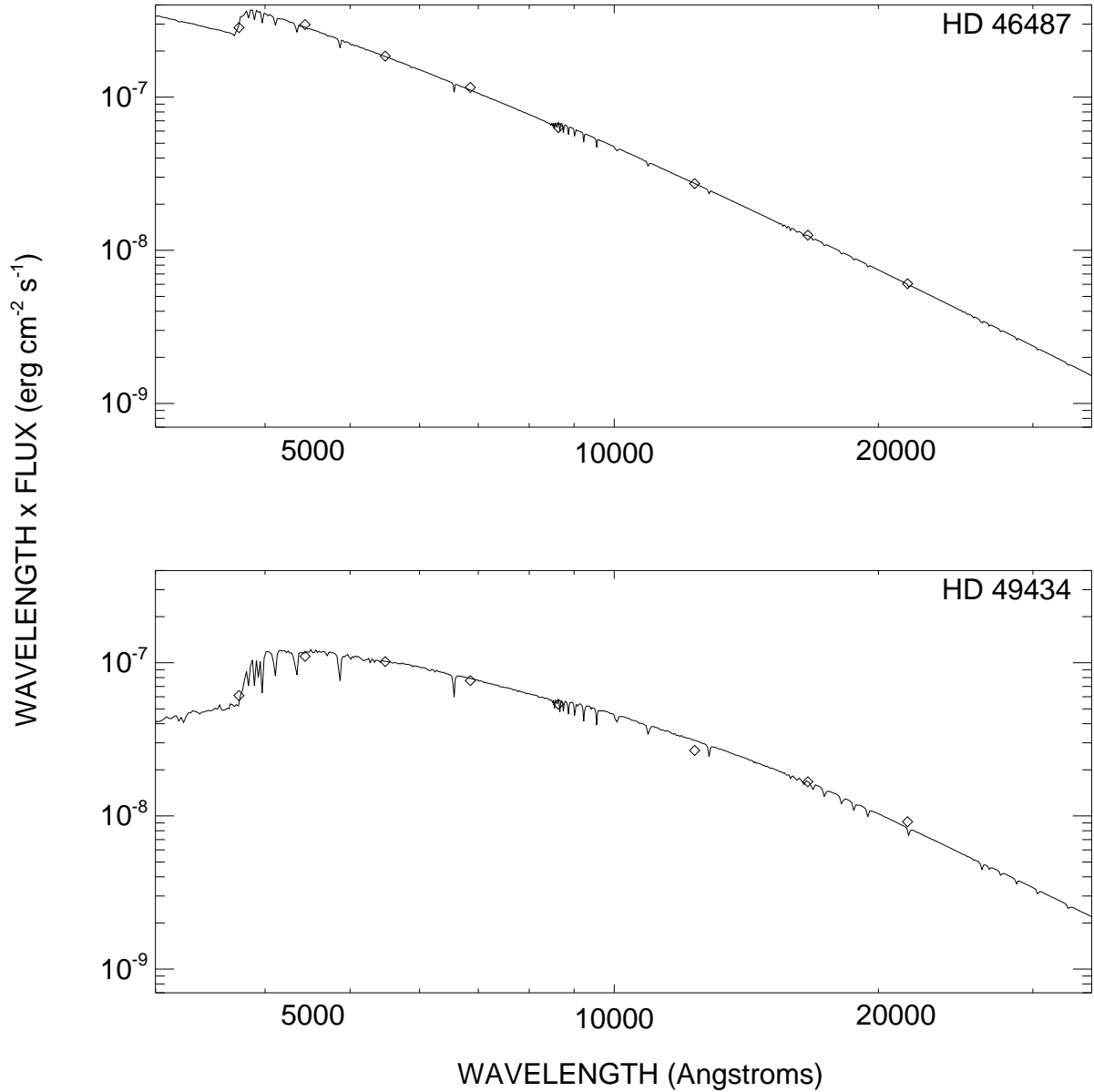}
\caption{SED fits for the calibrator stars. The diamonds are fluxes derived from $UBVRI JHK$ photometry (left to right) and the solid lines are the Kurucz stellar models of the stars with the best fit angular diameters. See Table \ref{calibrators} for the values used to create the fits.}
  \label{seds}
\end{figure}

We created spectral energy distribution (SED) fits to each calibrator star to check for possible unseen close companions. We used published $UBVRIJHK$ photometric values combined with Kurucz model atmospheres\footnote{Available to download at http://kurucz.cfa.harvard.edu.} based on $T_{\rm eff}$ and log~$g$ from the literature to estimate their angular diameters. The stellar models were fit to observed photometry after converting magnitudes to fluxes using \citet[][$UBVRI$]{1996AJ....112..307C} and \citet[][$JHK$]{2003AJ....126.1090C}. The photometry, $T_{\rm eff}$ and log~$g$ values, and resulting angular diameters for the calibrators are listed in Table \ref{calibrators}. There were no hints of excess emission associated with a low-mass stellar companion or circumstellar disk in the calibrators' SED fits (see Figure \ref{seds}).

\section{Results}
\subsection{Angular Diameter Measurement}
The observed quantity of an interferometer is defined as the visibility ($V$), which is fit with a model of a uniformly-illuminated disk (UD) that represents the observed face of the star. The diameter fit to $V$ was based upon the UD approximation given by $V = 2 J_1(x) / x$, where $J_1$ is the first-order Bessel function and $x = \pi B \theta_{\rm UD} \lambda^{-1}$, where $B$ is the projected baseline at the star's position, $\theta_{\rm UD}$ is the apparent UD angular diameter of the star, and $\lambda$ is the effective wavelength of the observation \citep{1992ARAandA..30..457S}. A more realistic model of a star's disk involves limb-darkening (LD), and the relationship incorporating the linear LD coefficient $\mu_{\lambda}$ \citep{1974MNRAS.167..475H} is:
\begin{equation}
V = \left( {1-\mu_\lambda \over 2} + {\mu_\lambda \over 3} \right)^{-1}
\times
\left[(1-\mu_\lambda) {J_1(x_{\rm LD}) \over x_{\rm LD}} + \mu_\lambda {\left( \frac{\pi}{2} \right)^{1/2} \frac{J_{3/2}(x_{\rm LD})}{x_{\rm LD}^{3/2}}} \right] 
\end{equation}
where $x_{\rm LD} = \pi B\theta_{\rm LD}\lambda^{-1}$.
Table \ref{calib_visy} lists the date of observation, the projected baseline $B$, the calibrated visibilities ($V$), and errors in $V$ ($\sigma V$).

\begin{figure}[h]
\includegraphics[width=0.75\textwidth, angle=90]{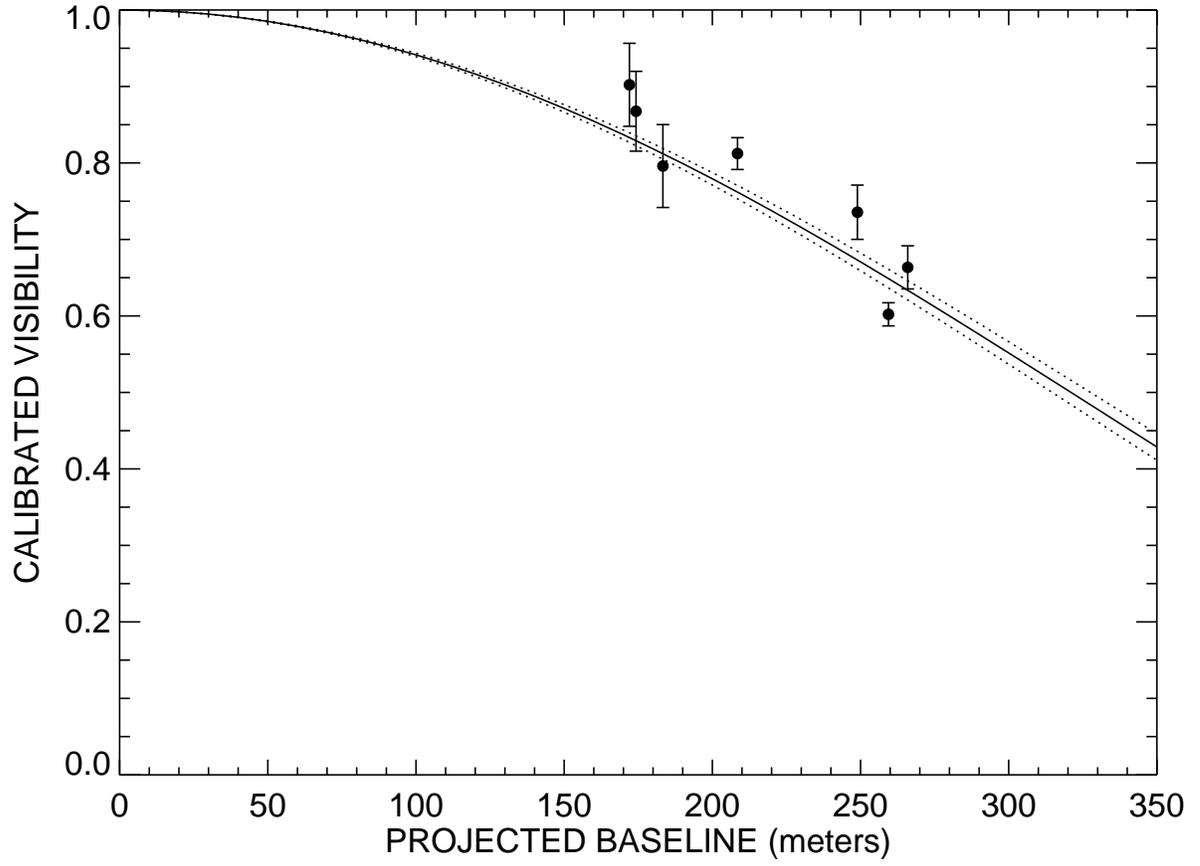}
\caption{HR 2582 $\theta_{\rm LD}$ fit. The solid line represents the theoretical visibility curve the best fit $\theta_{\rm LD}$, the dotted lines are the 1$\sigma$ error limits of the diameter fit, the filled circles are the calibrated visibilities, and the vertical lines are the measured errors.}
  \label{ldplot}
\end{figure}

The LD coefficient $\mu_{\rm K}$ of 0.31 was obtained from \citet{2011AandA...529A..75C} after adopting a $T_{\rm eff}$ of 4750 from \citet{2003AJ....125..359W} and a log $g$ of 2.14 cm s$^{\rm -2}$ from \citet{2000asqu.book.....C} for a K0 III, the spectral type listed in Wright et al. The resulting $\theta_{\rm UD}$ is 0.978$\pm$0.020 mas and $\theta_{\rm LD}$ is 1.005$\pm$0.020 mas, a 2$\%$ error. Figure \ref{ldplot} shows the $\theta_{\rm LD}$ fit for HR 2582. Limb-darkening is a second-order effect in the visibility curve that appears only after the first null in the visibility curve, i.e., when the visibility drops to zero. Because we are not beyond that null, we do not expect to see limb-darkening effects in our data and do not need to incorporate it into our model fit.

For the $\theta_{\rm LD}$ fit, the errors were derived using the reduced $\chi^2$ minimization method \citep{2003psa..book.....W,1992nrca.book.....P}: the diameter fit with the lowest $\chi^2$ was found and the corresponding diameter was the final $\theta_{\rm LD}$. The errors were calculated by finding the diameter at $\chi^2 + 1$ on either side of the minimum $\chi^2$ and determining the difference between the $\chi^2$ diameter and $\chi^2 +1$ diameter. The resulting $\chi^2$ is 19.8 and the reduced $\chi^2$ ($\chi_{\rm red}^2 =\chi^2$/DoF) is 3.3.\footnote{The degrees of freedom (DoF) is the number of observations minus the number of parameters fit to the data.} When the $\chi_{\rm red}^2$ is forced to be 1, $\chi^2$ is 6.0 and the errors nearly double from 0.020 to 0.036 mas. However, \citet{2010arXiv1009.2755A} describes why forcing $\chi_{\rm red}^2$ is not recommended: when the $\chi_{\rm red}^2$ is forced to be 1, it implies the model is completely correct, which is most often not the case. Even if the model is perfect, the DoF must be large in order to allow us to force the $\chi_{\rm red}^2$ to equal one with impunity. In this situation, our DoF is 6 so we use the errors associated with $\chi^2$, not $\chi_{\rm red}^2$.

\subsection{Stellar Radius, Luminosity and Effective Temperature}
HR 2582 has a parallax of 2.99$\pm$0.44 mas \citep{2007hnrr.book.....V}, which translates to a distance of 334.5$\pm$49.2 pc. When combined with our newly measured $\theta_{\rm LD}$, this gives us the physical radius of the star: 35.76$\pm$5.31 $R_\odot$, a error of 15$\%$. This is comparable to the radius determined by B12 of 34 $\pm$ 8 $R_\odot$ and provides better precision over their error of 24$\%$.

In order to determine the $L$ and $T_{\rm eff}$ of
HR 2582, we constructed its SED using photometric values published in \citet{1962MNSSA..21...20C}, \citet{1981PDAO...15..439M}, \citet{1987AandAS...68..259H}, \citet{1988iras....1.....B}, \citet{1972VA.....14...13G}, \citet{1991TrSht..63....1K}, \citet{1997AandAS..124..349M}, and \citet{2003tmc..book.....C}. The assigned uncertainties for the Two Micron All Sky Survey infrared measurements are as reported, and an error of 0.05 mag was assigned to the optical measurements. 

\begin{figure}[h]
\includegraphics[width=1.0\textwidth]{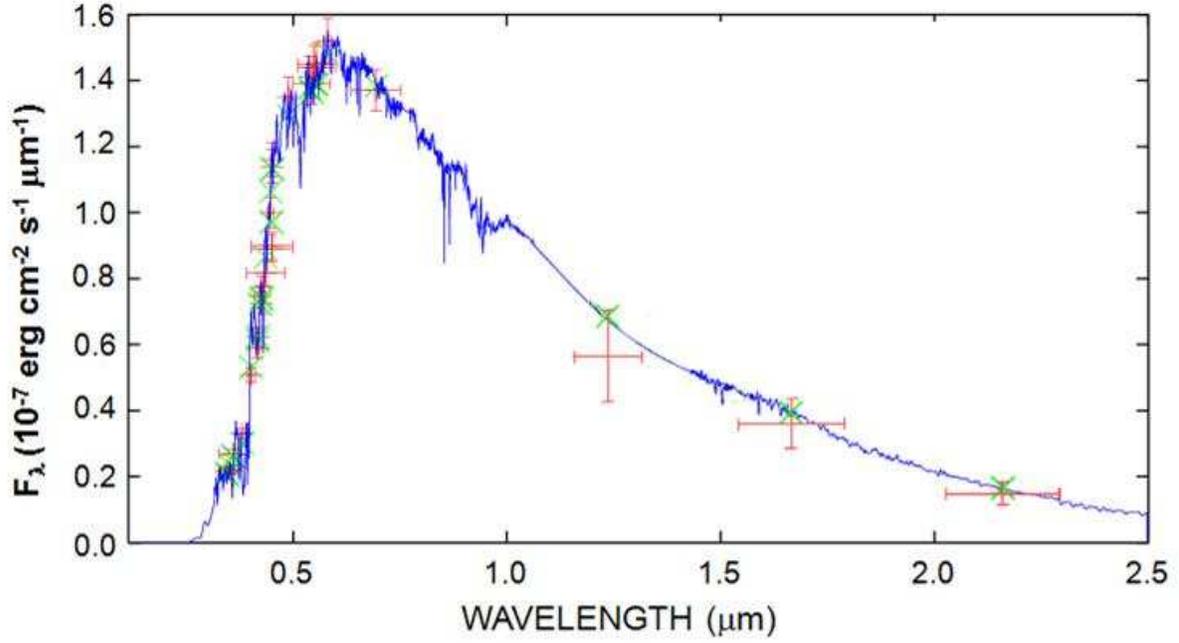}
\caption{HR 2582 SED fit. The solid-line spectrum is a K1 III spectral template from \citet{1998PASP..110..863P}. The crosses indicate photometry values from the literature and the horizontal bars represent bandwidths of the filters used. The X-shaped symbols show the flux value of the spectral template integrated over the filter transmission.}
  \label{sed}
\end{figure}

HR 2582's bolometric flux ($F_{\rm BOL}$) was determined by finding the best fit stellar spectral template from the flux-calibrated stellar spectral atlas of \citet{1998PASP..110..863P} using the $\chi^2$ minimization technique. This best SED fit allows for extinction, using the wavelength-dependent reddening relations of \citet{1989ApJ...345..245C}. The best fit was found using a K1 III template with an assigned temperature of $4656 \pm 120$ K, an extinction of $A_{\rm V}$ = 0.091 $\pm$ 0.042 mag, and a $F_{\rm BOL}$ of $1.48\pm0.05 \times 10^{-7} \mathrm{\; erg \; s^{-1} \; cm^{-2}}$. Figure \ref{sed} shows the best fit and the results are listed in Table \ref{results}.

We then combined $F_{\rm BOL}$ with HR 2582's distance to estimate its luminosity where $L = 4 \pi d^2 F_{\rm BOL}$, which produced a value of 517.8$\pm$ 17.5$L_\odot$. The uncertainty in $L$ is largely due to the uncertainty in the distance. The $F_{\rm BOL}$ was also combined with the star's $\theta_{\rm LD}$
to determine its effective temperature by inverting the relation
\begin{equation}
F_{\rm BOL} = {1 \over 4} \theta_{\rm LD}^2 \sigma T_{\rm eff}^4,
\end{equation}
where $\sigma$ is the Stefan-Boltzmann constant. This produces an effective temperature of $4579 \pm 60$ K, a 1$\%$ error. Because $\mu_{\rm K}$ is chosen based on a given $T_{\rm eff}$, we checked to see if $\mu_{\rm K}$ would change based on our new $T_{\rm eff}$ and iterated. $\mu_{\rm K}$ increased by 0.02 to 0.33, $\theta_{\rm LD}$ increased by only 0.001 mas to 1.006$\pm$0.020 mas, and $T_{\rm eff}$ decreased by 2 K to 4577 K, which are well within the errors. We adopted these $\theta_{\rm LD}$ and $T_{\rm eff}$ as our final values (see Table \ref{results}). The very slight change in $\theta_{\rm LD}$ did not affect the radius calculation. We also note the log $g$ used here (2.14 cm s$^{\rm -2}$) differs from that determined by B12 (1.4 cm s$^{\rm -2}$). We used B12's log $g$ to select $\mu_{\rm K}$ and the resulting change in $\mu_{\rm K}$ is +0.01 to 0.32 and we see above how little effect that has on the resulting $\theta_{\rm LD}$.

In Section 3.1, we compared the merits of $\chi^2$ versus $\chi^2_{\rm red}$ and leaned in favor of using $\chi^2$ errors. Those are the results listed in Table \ref{results}. However, if we do assume our model is perfect, force $\chi^2_{\rm red}$ to equal one, and use the resulting error of 0.036 mas in $\theta_{\rm LD}$, $\sigma_{T \rm EFF}$ increases from 60 to 91 K, an error of 2$\%$, and $\sigma_R$ remains the same at 5.31 $R_\odot$, an error of 15$\%$.

\section{Discussion}
As a check to our measurement, we estimated HR 2582's $\theta_{\rm LD}$ using two additional methods: (1) we used the SED fit as described in Section 3.2; and (2) we used the relationship between the star's dereddened ($V-K$) color \citep[calculated with the extinction curve described in][]{1989ApJ...345..245C}, $T_{\rm eff}$, and $\theta_{\rm LD}$ from \citet{1994AandA...282..899B}. Our measured $\theta_{\rm LD}$ is 1.006$\pm$0.020 mas, the SED fit predicts 0.972$\pm$0.053 mas, and the color-temperature-diameter relationship produces 0.926$\pm$0.369 mas. 

The main sources of errors for the three methods are uncertainties in visibilities for our interferometric measurement, uncertainties in the comparison between the observed and model fluxes for a given set of $T_{\rm eff}$ and log~$g$ values for the SED estimate, and uncertainties in the parameters of the relation and the spread of stars around that relation for the color-temperature-diameter determination. The three $\theta_{\rm LD}$ agree within the errors, and our interferometric measurements provide an error approximately 3 and 18 times smaller than the other methods, respectively.

\begin{figure}[h]
\includegraphics[width=1.0\textwidth]{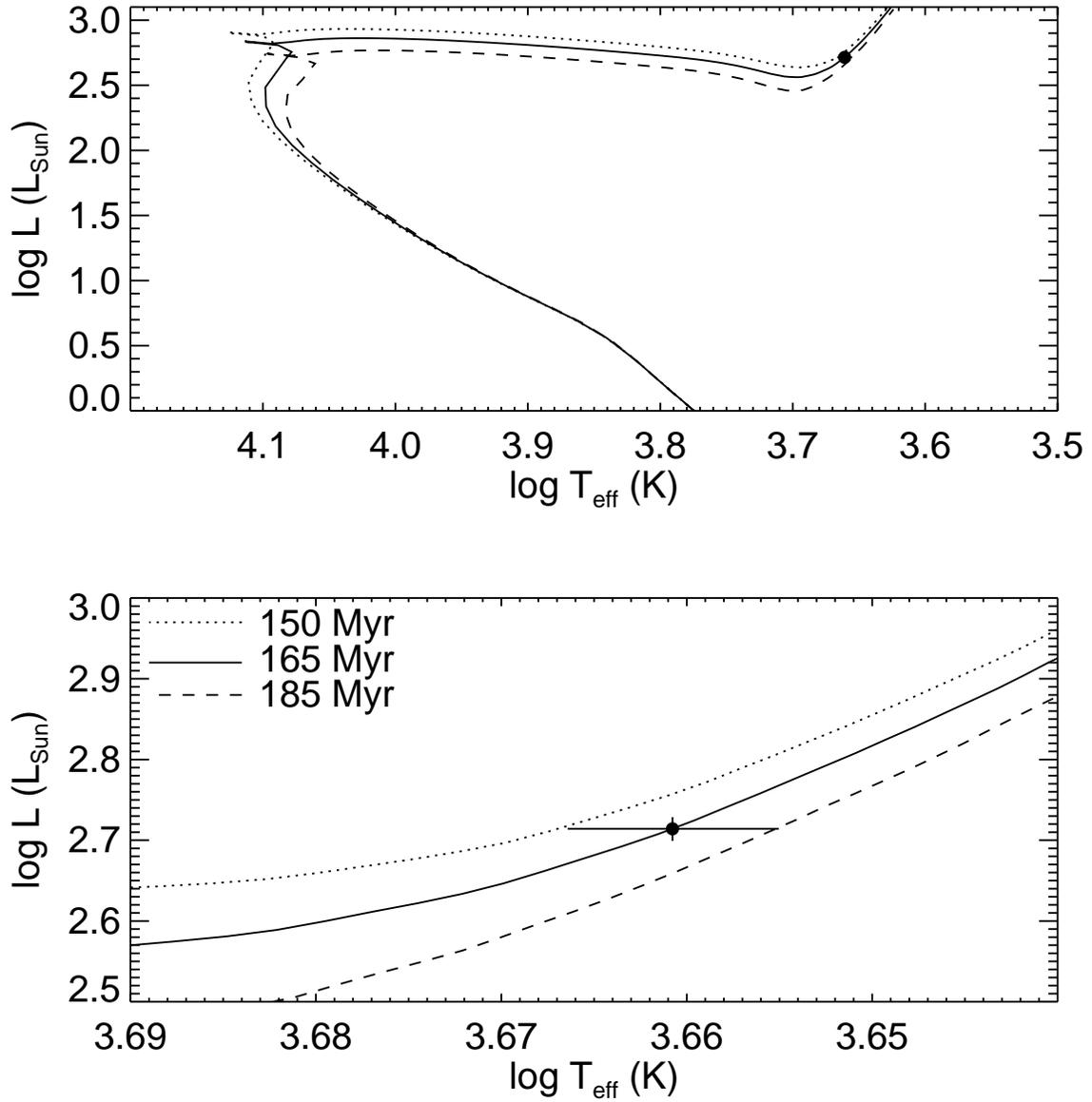}
\caption{H--R diagram for HR 2582. The lines are Y$^2$ isochrones for the ages indicated and the filled circle indicates our new $L$ and $T_{\rm eff}$ values with their associated errors. The bottom panel is a close-up of the top panel.}
  \label{yy}
\end{figure}

We used our new values of $T_{\rm eff}$ and $R$ in Equation 1 to calculate HR 2582's mass with the result of 5.6$\pm$1.7 $M_\odot$. This is slightly more massive than B12's mass of 5.2$\pm$2.9 $M_\odot$ but is well within the errors. We also used $T_{\rm eff}$ and $L$ to estimate the age of HR 2582 using the Yonsei--Yale isochrones \citep[$Y^2$,][]{2001ApJS..136..417Y}. We adopted [Fe/H] = --0.18 to be consistent with B12. The resulting age is 165$^{+20}_{-15}$ Myr (see Figure \ref{yy}), which is higher than the 105.5 Myr age quoted in B12. They do not discuss how they determined the age except to note that it is one of the model outputs, and do not give an error for that parameter.

This resulting age does not definitively answer the question of what evolutionary state HR 2582 is currently occupying. If the star is burning hydrogen in a shell on the first ascending branch, it is $\sim$157 Myr old. If it is burning helium in its core on the descending or second ascending branches, it is $\sim$163 Myr or $\sim$180 Myr old, respectively (B12). We lack the precision to determine exactly what is occurring in the interior of this star but it remains an excellent test case for stellar models, particularly with our more precise radius and temperature measurements.


\section{Summary}
We directly measured the limb-darkened angular diameter of HR 2582 with the CHARA Array interferometer and used our result of 1.006$\pm$0.020 mas along with the parallax measurement and photometry from the literature to calculate its physical radius (35.76$\pm$5.31 $R_\odot$), luminosity (517.8$\pm$17.5 $L_\odot$), and effective temperature (4577$\pm$60 K). We combined our $R$ and $T_{\rm eff}$ values with stellar oscillation results from B12 to determine the mass, which was 5.6$\pm$1.7 $M_\odot$ and the same $R$ and $T_{\rm eff}$ with $Y^{\rm 2}$ isochrones to estimate the star's age at 165$^{+20}_{-15}$ Myr.

\acknowledgments

The CHARA Array is funded by the National Science Foundation through NSF grant AST-0908253 and AST-1211129 and by Georgia State University through the College of Arts and Sciences, and by the W.M. Keck Foundation. STR acknowledges partial support by NASA grant NNH09AK731. This research has made use of the SIMBAD database, operated at CDS, Strasbourg, France.  This publication makes use of data products from the Two Micron All Sky Survey, which is a joint project of the University of Massachusetts and the Infrared Processing and Analysis Center/California Institute of Technology, funded by the National Aeronautics and Space Administration and the National Science Foundation.

\clearpage


\begin{deluxetable}{lcc}
\tablewidth{0pc}
\tablecaption{Calibrator Information.\label{calibrators}}
\tablehead{ \colhead{Parameter} & \colhead{HD 46487} & \colhead{HD 49434}}
\startdata
$U$ magnitude & 4.39 & 6.06 \\
$B$ magnitude & 4.95 & 6.03 \\
$V$ magnitude & 5.09 & 5.74 \\
$R$ magnitude & 5.14 & 5.59 \\
$I$ magnitude & 5.27 & 5.45 \\
$J$ magnitude & 5.38 & 5.40 \\
$H$ magnitude & 5.44 & 5.13 \\
$K$ magnitude & 5.46 & 5.01 \\
$E$($B-V$)      & 0.02 & 0.00 \\
$T_{\rm eff}$ (K) & 15200 & 7413  \\
log $g$ (cm s$^{-2}$) & 4.04 & 4.29 \\
$\theta_{\rm UD}$ (mas) & 0.210$\pm$0.004 & 0.347$\pm$0.016  \\
\enddata
\tablecomments{The photometric values are from the following sources: $UBV$ - \citet{Mermilliod}, $RI$ - \citet{2003AJ....125..984M}, $JHK$ - \citet{2003tmc..book.....C}. E($B-V$) was from \citet{1985ApJS...59..397S} for HD 46487 and \citet{2006AandA...458..293P} for HD 49434. $T_{\rm eff}$ and log $g$ was from \citet{2000asqu.book.....C} for HD 46487 based on its spectral type (B5 V) and from \citet{1999AandA...352..555A} for HD 49434. The uniform-disk angular diameters ($\theta_{\rm UD}$) are the result of the SED fitting procedure described in Section 2.}
\end{deluxetable}

\clearpage


\begin{deluxetable}{cccc}
\tablewidth{0pc}
\tablecaption{HR 2582 Calibrated Visibilities.\label{calib_visy}}

\tablehead{\colhead{MJD} & \colhead{$B$ (m)} & \colhead{$V$} & \colhead{$\sigma V$} \\ }
\startdata
56273.285 & 172.02 & 0.902 & 0.054 \\
56273.290 & 174.25 & 0.868 & 0.052 \\
56273.306 & 183.32 & 0.796 & 0.054 \\
56273.343 & 208.48 & 0.812 & 0.021 \\
56273.404 & 248.94 & 0.736 & 0.036 \\
56273.423 & 259.43 & 0.602 & 0.015 \\
56273.437 & 265.92 & 0.664 & 0.028 \\

\enddata
\end{deluxetable}

\clearpage

\begin{deluxetable}{lcl}
\tablewidth{0pc}
\tablecaption{HR 2582 Stellar Parameters.\label{results}}
\tablehead{ \colhead{Parameter} & \colhead{Value} & \colhead{Reference} }
\startdata
\cline{1-3}
\cline{1-3}
\multicolumn{3}{c}{From the literature:} \\
$V$ magnitude     & 6.04$\pm$0.01 & \citet{Mermilliod} \\
$K$ magnitude     & 3.65$\; \pm \;$0.28 & \citet{2003tmc..book.....C} \\
$\pi$ (mas) & 2.99$\; \pm \;$0.44 & \citet{2007hnrr.book.....V} \\
Distance (pc) & 334.5$\; \pm \;$49.2 & Calculated from $\pi$ \\
$\mu_{\lambda}$ & 0.33 & \citet{2011AandA...529A..75C} \\
\cline{1-3}
\cline{1-3}
\multicolumn{3}{c}{The results of our SED fit:} \\
$A_{\rm V}$       & 0.09$\; \pm \;$0.04 & \\
$F_{\rm BOL}$ (10$^{-7}$ erg s$^{-1}$ cm$^{-2}$) & 1.48$\; \pm \;$0.05 & \\
$T_{\rm eff,estimated}$ (K) & 4656$\; \pm \;$120 &  \\
$\theta_{\rm LD,estimated}$ (mas) & 0.972$\; \pm \;$0.053 &  \\
\cline{1-3}
\cline{1-3}
\multicolumn{3}{c}{The results of this work:} \\
$\theta_{\rm UD}$ (mas) & 0.978$\; \pm \;$0.020 & \\
$\theta_{\rm LD}$ (mas) & 1.006$\; \pm \;$0.020 &  \\
$R_{\rm linear}$ ($R_\odot$) &  35.76$\; \pm \;$5.31 &  \\
$T_{\rm eff}$ (K) & 4577$\; \pm \;$60 &  \\
$L$ ($L_\odot$) & 517.8$\; \pm \;$17.5  &  \\
Mass ($M_\odot$) & 5.6 $\pm$ 1.7 &  \\
Age (Myr) & 165$^{+20}_{-15}$ &  \\
\enddata
\end{deluxetable}

\clearpage


\clearpage

\clearpage

\clearpage

\end{document}